\documentclass[showpacs,prc]{revtex4}

\usepackage{graphicx}
\usepackage{dcolumn}
\usepackage{bm}
\usepackage{graphicx}
\usepackage{amssymb}
\usepackage{epstopdf}
\usepackage{fancybox}

\begin{document}
\title{\bf Electrostrong Nuclear Disintegration in Condensed Matter}
\author{J. Swain}
\affiliation{Physics Department, Northeastern University, Boston MA, USA}
\author{Y.N. Srivastava}
\affiliation{Department of Physics \& INFN, University of Perugia, Perugia, IT}
\author{A. Widom}
\affiliation{Physics Department, Northeastern University, Boston MA, USA}
\date{10 June 2013}                                           

\begin{abstract}

Photo- and electro-disintegration techniques have been traditionally used for studying giant dipole resonances and through them nuclear structure. Over a long period, detailed theoretical models for the giant dipole resonances were proposed and low energy electron accelerators were constructed to perform experiments to test their veracity. More recently, through laser and ``smart'' material devices, electrons have been accelerated in condensed matter systems up to several tens of MeV. We discuss here the possibility of inducing electro-disintegration of nuclei through such devices. It involves a synthesis of electromagnetic and strong forces in condensed matter via giant dipole resonances to give an effective ``electro-strong interaction''  - a large coupling of electromagnetic and strong interactions in the tens of MeV range.

\end{abstract}

\pacs{24.30.Cz,24.75.+i,25.20.-x,25.30.Rw,71.45.-d}

\maketitle

\section{Introduction}
Over several decades, virtual photons from electron scattering as well as Bremmstrahlung photons have been routinely used to cause nuclear photo-disintegration via the generation of giant dipole resonances (GDR) in the intermediate state\cite{Brink,Ishkhanov0}. The reactions studied extensively are with production of one or two neutrons such as $A(\tilde{\gamma}, n) A^*$ and $A(\tilde{\gamma}, nn) A^{**}$, where ($\tilde{\gamma}$) is the virtual photon from electron scattering and the final nuclei [$A^*, A^{**}$] stand for the disintegration product collectively. Of course, their counterpart nuclear breakup reactions $A(\gamma, n) A^*$ and $A(\gamma, nn) A^{**}$ from real Bremmstrahlung photons ($\gamma$) have also been of continued interest and study. Typically, GDR's are in the $(10 - 20) MeV$ range for heavy nuclei and $(15 - 25) MeV$ for light nuclei. Detailed compendia of such data exist\cite{Atlas} given their importance for a variety of practical applications\cite{Liu,Mao}.  

In the above mentioned type of accelerator experiments, it is rather hard to study directly the final nuclei produced [$A^*, A^{**}$] given their very low velocity and hence an exceedingly low probability for their escape. The object here is complementary. In section II we give a brief review of giant dipole resonances and their photoexcitation. Section III discusses GDR
excitation by electrons and the decays of GDR-excited nuclei into final states with neutrons. Section IV reviews some of the applications of GDR's, making it clear that the physics is
very well established both within nuclear physics proper and in a much wider scope. In section V we raise the possibility that if, in some condensed matter systems, high energy electrons (several MeV) are produced, then electrostrong photodisintegration may occur. In section VI we point out that if such processes occur, they should be accompanied
by the production of $e^+e^-$ pairs, and the annihilation of positrons produced in this way would be clear evidence of high energy electrons being produced. Section VII discusses
electrostrong induced endothermic fission which can take place in addition to the more common exothermic fission and alterations in exothermic fission rates as well as other
transmutations which can occur. Section VIII discusses a particularly interesting example in which aluminum and silicon might appear in an initial sample of iron. Section IX points out
that at scales of a few tens of MeV, electromagnetic, weak and strong interactions are all expected to occur in condensed matter and in section X we make our final conclusions.

\section{Photoexcitation of Giant Dipole Resonances}

Giant dipole resonances (GDR's) are collective excitations of atomic nuclei and can be thought of as collective oscillation of
protons against neutrons. Originally observed by Baldwin and Klaiber\cite{BK} as a strong absorption of photons around 15-20
MeV by nuclei, they were interpreted as collective proton oscillations to form a dipole by Goldhaber and Teller a year later\cite{GT} 
in 1948. GDR's are present in all nuclei and are a well-established phenomenon discussed in most nuclear physics
textbooks\cite{Blatt-W}, even if only briefly in more recent ones\cite{Walecka}. Following Chomaz\cite{Chomaz} we summarize
the basic properties of giant dipole resonances.
For spherical nuclei, the cross section for photoabsorption cross section $\sigma_{abs}(E_\gamma)$ as a function of photon
energy $E_\gamma$ can be approximated 
by

\begin{equation}
\sigma_{abs}(E_\gamma) = \frac{\sigma_{GDR}E_\gamma^2\Gamma_{GDR}^2}{(E_\gamma^2-{E_{GDR}^2})^2+E_\gamma^2\Gamma_{GDR}^2}
\end{equation}

\noindent where $E_{GDR}$ and $\Gamma_{GDR}$ are the energy and width of the resonance and $\sigma_{GDR}$ is its
maximum. For deformed nuclei there can be different dipole resonances corresponding to directions along and perpendicular
to the nuclear symmetry axis, a point we mention for completeness, but which will not be important in what follows.

The dependence of $E_{GDR}$ on 
nucleon number A can be parametrized by 

\begin{equation}
E_{GDR}=31.2A^{-1/3}+20.6A^{-1/6} (MeV)
\end{equation}

\noindent or, in the case of medium and heavy nuclei, more succinctly as 

\begin{equation}
E_{GDR}\approx 80 A^{-1/3} MeV
\end{equation}

The width $\Gamma_{GDR}$ is typically between 4 and 8 MeV with the narrower widths being seen for
magic nuclei (nuclei with neutrons or protons forming complete shells with 2, 8, 20, 28, 50, 82, or 126 particles).

An approximate expression for the dipole strength integrated up to $E_\gamma=30$ MeV is given by the Thomas
Reiche-Kuhn sum rule \cite{Le50}

\begin{equation}
m_1=\frac{2\pi e^2h}{mc}\frac{NZ}{A}\sim 60 \frac{NZ}{A} MeV \cdot mb
\end{equation}

Ejiri and Dat\'{e}\cite{Ejiri-Date} have recently provided a very useful  
approximate formula for the cross section at the peak in terms of 
nucleon number $A$ of

\begin{equation}
\sigma_{GDR}\approx (3\times 10^{-3} A) {\mathrm{barns}}.
\end{equation}

\noindent 
It is also worthy of note
that 
the peak value of the GDR cross section is typically a few percent of the pair production and
Compton scattering cross sections.

We summarize the important point we want the reader to take from this section as:

\medskip

\doublebox{\begin{minipage}{6.8in}Giant dipole resonances are well-studied and represent a strong coupling between
all atomic nuclei and photons in the range of 10-25 MeV.\end{minipage}}

\medskip

\section{Excitation of Giant Dipole Resonances by Electrons and Decays of Giant Dipole Resonances}

The photon exciting a nuclear giant dipole resonance need not be on-shell, and it is well-known that
electrons can excite these resonances in bulk matter either through Bremmstrahlung of an on-shell photon
due to an electron passing by other nuclei (in which the above formulae apply) or via direct exchange of a virtual photon with
a nucleus. Again, there is a vast literature. For example, Barber and George \cite{Barber-George} measured neutron
yields from C, Al, Cu, Ta, Pb and U targets with electrons from 10 to 36 MeV, finding, for thick targets, yields of
around $10^{-3}-10^{-2}$ from the decays of giant dipole resonances. 

Giant dipole resonances predominantly decay by the emission of one or two neutrons, possibly followed by
fission of the remaining nucleus. Detailed calculations of photo
and electroproduction rates 
for producing neutrons
in a variety of materials have been done\cite{Liu,Mao} due to the interest in activation of materials at medical accelerators working
in the range of a few tens of MeV. Both slow ($<2.5$ MeV) neutrons from evaporation and a smaller fraction of faster neutrons
can be produced.

Mao {\em et al.} \cite{Mao} give an approximate formula for the neutron yield per electron per MeV of electron energy
via giant dipole resonances as $8 \times 10^{-6} (Z^{1/2}+0.12 Z^{3/2}-0.001 Z^{5/2})$ for thick targets.

We summarize the important point we want the reader to take from this section as:

\medskip

\doublebox{\begin{minipage}{6.8in}Giant dipole resonances are well-known to be excited by electrons with a few tens
of MeV with significant neutron yields (often $10^{-3}$ or more) per electron on thick targets, and both fast and slow
neutrons can be produced. \end{minipage}}

\medskip

\section{Applications of Giant Dipole Resonances}

We want to emphasize that giant dipole resonances are very well-known across many disciplines beyond nuclear physics
proper as discussed above. For example, giant dipole resonances mediate
the energy loss of high energy nuclei due to dissociation on the cosmic microwave background \cite{Stanev}.
Giant dipole resonances are also well-known in astrophysical nucleosynthesis \cite{Ishkhanov}. Ejiri and Dat\'{e}\cite{Ejiri-Date}
have proposed Compton-backscattered laser photons from GeV electrons for the production of useful radioactive isotopes
(for example, for medical applications) via giant dipole resonances.
It has also been suggested that radioactive waste products such as $^{129}I$ could be transmuted via electron-beam
induced giant dipole resonances and their subsequent decays, with transmutations (on another isotope for safety). These have
actually been carried\cite{Li} out at NewSUBARU
in Japan using 1064 nm laser photons from a Nd:YVO laser, Compton scattered from a stored electron beam to energies
up to 17.6 MeV. Iodine-129 has been transmuted using a laser-generated plasma to accelerate electrons to produce
gamma rays to excite the giant dipole resonance\cite{laser-I1,laser-I2}. Giant dipole resonances, driven by lasers, have
also been used as sources of intense gamma ray sources\cite{gamma-source}. For a very comprehensive review of 
laser-driven nuclear processes, see \cite{Lasers-and-nuclei}.

We summarize the important point we want the reader to take from this section as:

\medskip

\doublebox{\begin{minipage}{6.8in}Giant dipole resonances are very well understood and used, both theoretically and practically in devices
well outside the scope of nuclear physics proper. \end{minipage}}

\medskip

\section{The Possibility of Induced Nuclear Disintegration in Condensed Matter}

With it now clear that the phenomenon of giant dipole resonances is well-established and expected if electrons of 10-25 MeV
are present, we now ask whether or not such phenomena might take place in condensed matter without the use of a traditional
electron beam from an accelerator or a similar photon beam (for example, from Compton backscattering).

The answer is already affirmative in the cases cited above of electrons driven to high energies by the direct action of low energy (eV scale)
laser photons on bulk condensed matter.

In recent years it has been suggested that electrons may be accelerated to several tens of MeV in condensed matter systems
including, but not limited to, electrochemical cells \cite{Widom-electroweak} and the fracture of piezoelectric materials such
as rocks\cite{piezo-jphysg} with associated neutron production via $e+p\rightarrow n+\nu_e$ calculable within 
the usual Fermi theory of
the weak interaction (or, more completely, by the Standard Model). For general reviews which also introduce other
systems such as exploding wires where collective effects can produce such large electron energies, see \cite{LENR1,LENR2}.

We now suggest that {\em if} electrons reach such energies in these processes {\em then} there is an additional testable
consequence, which is the appearance of nuclear reactions mediated by giant dipole resonances. Indeed, these processes, being unsuppressed by
powers of the Fermi constant, would be expected to occur more readily and could provide not only neutrons but also fission products
from the decays of GDR excited nuclei as well as new nuclei formed from the absorption of produced neutrons and, where applicable,
their subsequent decay. 

The central message from this section is:

\medskip

\doublebox{\begin{minipage}{6.8in} {\em If} electrons are accelerated to tens of MeV in condensed matter systems, {\em then} in addition
to producing neutrons via electroweak processes, one expects, and at much higher rates, what we call ``electrostrong processes'',
where nuclear reactions take place mediated by giant dipole resonances. In this case one expects slow neutrons from evaporation of
GDR's as well as some fast ones, and additional nuclear reactions when those neutrons are absorbed. \end{minipage}}

\medskip

\section{Electron-Positron Pair Production}

In addition, as noted earlier \cite{Ejiri-Date}, at even greater rates, one should see electron-positron pair production. Positrons produced
in such processes would be expected to stop and annihilate on electrons giving the well-known signature of two back-to-back 511 keV
photons. We suggest looking for such gamma ray lines to investigate the production of antimatter where electrons reach energies sufficient
to excite giant dipole resonances.

The central message from this section is:

\medskip

\doublebox{\begin{minipage}{6.8in} {\em If} electrons are accelerated to tens of MeV in condensed matter systems, {\em then} 
electron-positron pair creation would also be expected at comparable (higher) rates. \end{minipage}}

\medskip

\section{Endothermic fission and other transmutations}

Fission is often only considered for nuclei heavier than iron, which sits at the bottom of the binding energy
curve - anything lighter than iron needs energy to be supplied to get it to split into lighter nuclei. Very little seems to have
been done by way of analysing the decay products of GDR-excited nuclei other than to count the neutrons released.
At higher energies (1.5 GeV and over) one sees a wide variety of decay products \cite{Fulmer} but if one just has
enough energy to excite the GDR, there seems to be little known.

Now if tens of MeV are present in simple condensed matter systems, and with the giant dipole resonances available, endothermic
fission reactions may be more interesting and more common than has been typically thought. Looking for new elements or
new isotopes not present originally would indicate the occurrence of nuclear reactions in addition to the simple detection of neutrons
(many of which may be too slow to make it to detectors, but which could reveal themselves through transmutations). We emphasize
that since the processes considered here, unlike earlier electroweak low energy nuclear reactions, are not suppressed by the Fermi
constant, the scale at which transmutations occur could be very large - on the order of $10^{-3}$ or more per energetic electron
as discussed above.

Of course one can also expect increased rates for exothermic fission reactions, such as increased rates of spontaneous nuclear
fission processes. Whatever nuclei are produced, they may in turn undergo further reactions such as decays (weak or strong, or 
through emission of gamma rays) and may absorb neutrons such as those produced in the initial GDR decay.

Obviously this idea opens up a vast range of possibilities to consider, with searches for new nuclei not originally present. These might
be revealed via chemical means, or via neutron activation, electron microscopy elemental analysis, X-ray fluoresence, or other techniques.
In the next section, for interest, and because there may be experimental data to  support it\cite{Carpinteri}, we consider a simple example.

The central message from this section is:

\medskip

\doublebox{\begin{minipage}{6.8in} {\em If} electrons are accelerated to tens of MeV in condensed matter systems, {\em then} 
one expects both endothermic and exothermic nuclear fission processes as well as the appearance of new nuclei due
to further reactions of the decay products including further decays and/or the absorption of produced neutrons.\end{minipage}}

\medskip

\section{An example: Aluminum and Silicon from Iron}

Motivated by \cite{Carpinteri}, consider the processes  $^{56}_{26}Fe +\gamma \rightarrow 2 [ ^{28}_{14}Si  + e^- + \bar{\nu}_e]$
and  $^{56}_{26}Fe +\gamma \rightarrow 2 [ ^{27}_{13}Al  +  n]$.\cite{Widom}. 
 
 Upon absorbing a photon, the excited nucleus $^{56}_{26}Fe^*$ may fission symmetrically into two identical $^{28}_{13}Al$ nuclei.
The nucleus $^{28}_{13}Al$ $\beta^-$ decays into a stable $^{28}_{14}Si$ nucleus and an $e^-$ plus a $\bar{\nu}_e$ in about $2$ 
minutes. From the point of view of iso-spin, the iron nucleus had $4$ extra neutrons $I_3 = -2$ with each pair of neutrons being in  
relative orbital angular momentum $L=0$; spin $S=0$ (antisymmetric) and isospin $I =1$ (symmetric state). Each of the weak decay 
carries away $I =1$ thus leading to $^{28}_{14}Si$ in the $I=0$ state.

On the other hand, there is also the possibility that two neutrons come out and the iron nucleus breaks up into two identical stable 
$^{27}_{13}Al$ nuclei [along with the two free neutrons of course] as alluded to above. The minimum photon energy to allow this is 
42.350354 MeV. Each pair [$^{27}_{13}Al + n$] would be in an iso-spin $I =1$ state.  

The neutrons produced could be absorbed by other $^{56}Fe$ (which is the main stable isotope, making up almost 92\% of naturally occurring
iron) to produce more $^{57}Fe$ and $^{58}Fe$, both of which are stable and might be detectable if produced in significant amounts.

Neutrons could also be absorbed by aluminum \cite{Aluminum-activation}.
If neutrons are absorbed by aluminum, silicon can be produced via $^{27}Al + n \rightarrow ^{28}Al$
(which has a thermal neutron capture cross section of ~12 barns) with the $^{28}Al$ undergoing beta decay to $^{28}Si$. Other transmutation reactions
include $^{27}Al (n,p)^{27}Mg$, $^{27}Al (n,\alpha)^{24}Na$ which beta decay to excited states of  $^{24}Mg$ and $^{27}Al$ respectively
which then decay by gamma emission. All the beta and gamma emissions could be looked for in addition to direct searches for new elements.

\medskip   
%
%

The central message from this section is:

\medskip

\doublebox{\begin{minipage}{6.8in} {\em If} electrons are accelerated to several tens of MeV in condensed matter systems containing iron, {\em then} 
one may expect the appearance of aluminum and silicon.\end{minipage}}

\medskip

\section{An Approximate Unification of Forces at $\sim$ 10-20 MeV in Condensed Matter}

It is an old idea \cite{unif} that the coupling constants for the electromagnetic, weak, and strong interactions
- coupling constants for $U(1)$, $SU(2)$ and $SU(3)$ gauge theories - might run with energy via the renormalization
group such that they would take on the same value at some very large energy scale, presumably of the order of $10^{16}$GeV.

Here we find it amusing to note that it is within the energy range of ~(10-20) MeV that electromagnetic,
weak, and strong processes all start to become possible in condensed matter. That is, if one imagines a developing civilization
familiar with electromagnetism, trying to go to higher and higher energies and with no prior direct knowledge of the
weak and strong interactions, they would discover them at about the same energy scale.

The central message from this section is:

\medskip

\doublebox{\begin{minipage}{6.8in} At tens of MeV, electromagnetic, weak, and strong processes can all be expected
to occur in bulk condensed matter.\end{minipage}}

\medskip

\section{Conclusions}

Collective phenomena in nuclei, via the 
ubiquitous giant dipole resonances, provide strong couplings
between electromagnetism and the strong interaction at a scale of a few tens of MeV and well below the mass of
the pion, which might have been thought of naively as a sort of threshold. Similarly, it has been suggested that collective
phenomena in condensed matter can give rise to electrons of a similar energy scale. If these suggestions are 
indeed realizable, then
a wide range of nuclear phenomena (as well as $e^+e^-$ pair creation) should occur with measurable rates
and novel consequences.

\section{Acknowledgments}

We would like to thank A. Carpinteri, M. Manuello, G. Lacidogna, O. Borla and their colleagues at Politecnico di Torino 
for some very interesting discussions and for sharing with us some of their preliminary experimental results. J. S. 
would like to thank the National Science Foundation for support under grant PHY-1205845.

\end{document}